\documentclass[twocolumn,aps,prl,floatfix,showpacs,shownokeys,epsfig,grapics]{revtex4}%
\usepackage{graphicx}
\usepackage{amsmath}
\usepackage{amsfonts}
\usepackage{amssymb}
\newcommand{\ket}[1]{|{#1} \rangle}
\newcommand{\bra}[1]{\langle {#1}|}

\newcommand{\bk}{{\mathbf k}}

\newcommand{\br}{{\mathbf r}}

\newcommand{\up}{{\uparrow}}
\newcommand{\down}{\downarrow}

\newcommand{\cd}{c^{\dag}}
\begin{document}
\title{Entanglement of an impurity and conduction spins in the Kondo model}
\author{Sangchul Oh}\email{scoh@kias.re.kr}
\author{Jaewan Kim}\email{jaewan@kias.re.kr}
\affiliation{School of Computational Sciences, 
             Korea Institute for Advanced Study, Seoul 130-722, Korea}
\date{\today}
\begin{abstract}
Based on Yosida's ground state of the single-impurity Kondo Hamiltonian, 
we study three kinds of entanglement between an impurity and conduction 
electron spins. First, it is shown that the impurity spin is maximally 
entangled with all the conduction electrons. Second, a two-spin density 
matrix of the impurity spin and one conduction electron spin is given by 
a Werner state. We find that the impurity spin is not entangled with one 
conduction electron spin even within the Kondo screening length $\xi_K$, 
although there is the spin-spin correlation between them. Third, we show 
the density matrix of two conduction electron spins is nearly same to that 
of a free electron gas. The single impurity does not change the entanglement 
structure of the conduction electrons in contrast to the dramatic change 
in electrical resistance.
\end{abstract}
\maketitle

Entanglement, the quantum correlation between subsystems, is considered 
to be one of the key concepts in quantum mechanics and quantum information 
science~\cite{Nielsen00}. A study of entangled structures of quantum 
many-body systems is of great importance for providing us not only a new view 
on their physical properties but also the basic knowledge for fabricating 
quantum information processors. For example, entanglement of the ground 
states of one-dimensional quantum spin lattice models has been intensively 
investigated in context of quantum phase transition~\cite{Osborne02,Vidal03}.
In a non-interacting electron gas, entanglement of two electron spins due to 
the Pauli exclusion principle has been studied in connection with its
characteristic length scale, the Fermi wavelength 
$\lambda_F$~\cite{Vedral03,Oh04,Lunkes05}. Entanglement of two electron 
spins of a BCS superconductor has also examined with relation to
the coherence length $\xi_0$~\cite{Oh05}.

\begin{figure}
\includegraphics[scale=0.7,angle=0]{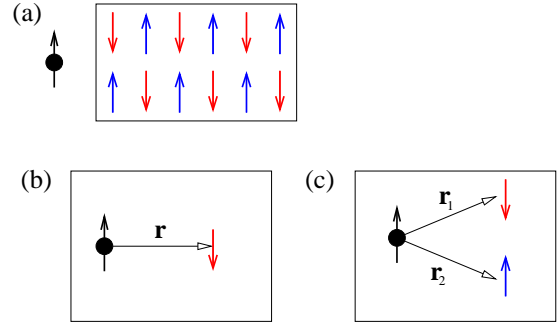}
\caption{(color online). Schematic of three kinds of entanglement: 
entanglement (a) between the impurity spin and all the conduction electrons, 
(b) between the impurity spin and one conduction electron spin at $\br$, 
and (c) between two conduction electrons at $\br_1$ and $\br_2$. 
The rectangular box is the conduction electrons, the arrow 
with a filled circle represents the impurity spin, and the red and blue
arrows denote the conduction electron spins.}
\label{Fig:fig1}
\end{figure}

The Kondo model describing the exchange interaction between the impurity spin 
and the conduction electrons has been one of challenging quantum many-body 
problems in condensed matter physics~\cite{Hewson93,Kouwenhoven01,Yosida96}. 
Various theoretical tools such as Anderson's scaling theory, Wilson's numerical 
renormalization group approach, and the Bethe ansatz, etc. have been applied 
to the Kondo model~\cite{Hewson93}. Recently, Kondo effects in nanodevices 
have been revisited with potential applications to spin devices or spin 
qubits~\cite{Kouwenhoven01,Craig04,Glazman04}. Although the Kondo model have 
been very intensively studied, its entanglement structure remains 
to be explored. In this paper we investigate three kinds of entanglement 
between the impurity spin and conduction electron spins as shown 
in Fig.~\ref{Fig:fig1}. From this study on the entanglement structure of 
the Kondo model we provide a clear view on the Kondo screening cloud 
with the size of the Kondo screening length $\xi_K$, which is the holy grail
in the Kondo physics ~\cite{Kouwenhoven01,Soerensen96}.
Also we discuss a possibility for the use of an electron gas in coupling 
spin qubits~\cite{Craig04,Glazman04}.

Let us consider the spin-$\frac{1}{2}$ Kondo Hamiltonian 
\begin{align}
H = \sum_{\bk\sigma}\epsilon_{\bk} \cd_{\bk\sigma}c_{\bk\sigma}
  - 2J{\mathbf S}\cdot{\mathbf s}(0) \,,
\label{Eq:Hamil}
\end{align}
where ${\mathbf S}$ is the spin of the magnetic impurity, ${\mathbf s}(0)$ 
the spin of the conduction electrons at $\br = 0$, and $\cd_{\bk\sigma}$ 
creates a conduction electron with momentum $\bk$ and spin $\sigma$. 
The exchange interaction $J$ is assumed to be negative, so the impurity spin 
is anti-parallel to the conduction spins.

Yosida presented the variational ground state of Eq.~(\ref{Eq:Hamil}) given 
by a singlet state~\cite{Hewson93,Yosida66,Yosida96,Yamada04} 
\begin{align}
\ket{\Psi_s} 
= \frac{1}{\sqrt{2}} \left(\, 
  \ket{\phi_\down} \ket{\chi_\up} -\ket{\phi_\up}\ket{\chi_\down} 
                   \,\right) \,,
\label{Eq:singlet}
\end{align}
where $\ket{\chi_\up}$ is the spin-up state of the impurity spin and 
$\ket{\phi_\down}$ denotes the state of conduction electrons with one 
excess down spin. The simplest form of $\ket{\phi_\down}$ is the state 
with one down spin added outside the filled Fermi sphere 
\begin{align}
\ket{\phi_\down} = \frac{1}{\sqrt{\cal{N}}}
\sum_{k>k_F} \Gamma_{\bk} \cd_{\bk\down} \ket{F}\,,
\end{align}
where $\Gamma_{\bk}$ are variational parameters, 
${\cal N}$ the normalization factor, and 
$\ket{F} = \prod_{k\le k_F} \cd_{\bk\sigma}\ket{0}$ the filled Fermi sphere.
The variational parameters $\Gamma_{\bk}$ are given by
\begin{align}
\Gamma_{\bk} = \frac{1}{\epsilon_{\bk} + E_B}\,,
\end{align}
where $E_B = k_BT_K = D\, \exp[2/3JN(0)]$. Here $T_K$ is called the Kondo 
temperature, $2D$ the bandwidth of the conduction electrons, 
and $N(0)$ the density of states of the conduction 
electrons. The characteristic length scale of the Kondo model is 
the Kondo screening length $\xi_K$, the size of the Kondo screening cloud
\begin{align}
\xi_K = \frac{\hbar v_F}{k_BT_K} = \frac{E_F}{k_BT_K} \frac{2}{k_F}\,,
\label{Eq:screening}
\end{align}
where $v_F$ is the Fermi velocity, $E_F$ the Fermi energy, and $k_F$ 
the Fermi momentum. Since $E_F \gg k_BT_K$, $\xi_K$ is much larger 
than the Fermi wavelength $\lambda_F=2\pi/k_F$. 

\paragraph{(i) Entanglement between the impurity spin and the conduction 
electrons.} As depicted in  Fig.~\ref{Fig:fig1}-(a), let us investigate 
entanglement between the impurity spin and all the conduction electrons.
Since the total state, 
Eq.~(\ref{Eq:singlet}) is a pure state, entanglement between them is 
quantified by the von Neumann entropy of the reduced density matrix 
$\rho_{\rm im}$ of the impurity spin,
$S(\rho_{\rm im}) = -{\rm Tr}\left[\rho_{\rm im}\log\rho_{\rm im}\right]$.
Here the subscript `im' represents the impurity. 
By tracing out the degrees of freedom of the conduction electrons in
Eq.~(\ref{Eq:singlet}), one easily obtains $\rho_{\rm im} 
= {\rm Tr}_{\rm con} \ket{\Psi_s}\bra{\Psi_s}$. 
It is given by the fully mixed state
\begin{align}
\rho_{\rm im} 
= \frac{1}{2} \begin{bmatrix}
               1 & 0 \\
               0 & 1
               \end{bmatrix}\,,
\end{align}
which gives the maximum entropy $S(\rho_{\rm im}) = 1$. Thus we find that 
the impurity spin is maximally entangled with the conduction electrons.

\paragraph{(ii) Entanglement of the impurity spin and one conduction electron.}
If a magnetic impurity is immersed in a Fermi sea, conduction electrons 
will align in order to screen the magnetic field by the impurity. 
To put it simply, this magnetic disturbance is called the Kondo screening cloud 
of electrons (or an electron) with the size of order $\xi_K$, forming 
a singlet with an impurity~\cite{Soerensen96}. So one may expect some 
correlation or entanglement between the impurity spin and a conduction 
electron in the Kondo screening cloud.

Let us consider the impurity spin and one conduction spin at $\br$ from 
the impurity as shown in Fig.~\ref{Fig:fig1}-(b). Entanglement between 
them can be measured if one obtains the two-spin density matrix by tracing out 
all the degrees of freedom of Eq.~(\ref{Eq:singlet}) except the impurity spin and 
one conduction spin at $\br$. In the second quantization, it is given as follows
\begin{align}
\rho^{(2)}_{\alpha\beta;\alpha'\beta'}(\br)
= \frac{1}{2} \bra{\Psi_s} 
  \hat{\psi}^{\dag}_{\beta'}(\br) \hat{\phi}^\dag_{\alpha'}(0) 
  \hat{\phi}_{\alpha}(0)\hat{\psi}_{\beta}(\br)
              \ket{\Psi_s}\,,
\label{Eq:tsdm_def}
\end{align}
where an operator $\hat{\psi}^{\dag}_{\beta}(\br)$ creates a conduction electron with
spin $\beta$ at $\br$ and $\hat{\phi}_{\alpha}$ is a creation operator of the impurity
spin $\alpha$ at origin. Here $\alpha,\alpha',\beta,\beta'$ refer to the spin indices,
i.e., $\up$ or $\down$.

From Eqs.~(\ref{Eq:singlet}) and (\ref{Eq:tsdm_def}), we obtain
\begin{align}
\rho^{(2)}(\br) 
= \frac{1}{2}
 \begin{bmatrix}
  \rho_{\up\up;\up\up} &0 & 0 & 0 \\[6pt]
  0 &\rho_{\up\down;\up\down} & \rho_{\up\down;\down\up} & 0 \\[6pt]
  0 &\rho_{\down\up;\up\down}  &\rho_{\down\up;\down\up} & 0 \\[6pt]
  0 & 0 & 0 &\rho_{\down\down;\down\down}
  \end{bmatrix} \,,
\end{align}
where the diagonal elements are given by 
\begin{subequations}
\begin{align}
\rho_{\up\up;\up\up}
  &= \bra{\phi_\down} \hat{\psi}^{\dag}_\up(\br) \hat{\psi}_{\up}(\br) 
     \ket{\phi_\down} \,,\\
\rho_{\up\down;\up\down} 
  &= \bra{\phi_\down} \hat{\psi}^{\dag}_\down(\br) \hat{\psi}_{\down}(\br) 
     \ket{\phi_\down} \,,\\
\rho_{\down\up;\down\up} 
  &= \bra{\phi_\up} \hat{\psi}^{\dag}_\up(\br) \hat{\psi}_{\up}(\br) 
     \ket{\phi_\up} \,,\\
\rho_{\down\down;\down\down} 
  &= \bra{\phi_\up} \hat{\psi}^{\dag}_{\down}(\br) \hat{\psi}_{\down}(\br) 
     \ket{\phi_\up} \,,
\end{align}
and the off-diagonal elements are given by 
\begin{align}
\rho_{\up\down;\down\up} 
= - \bra{\phi_\down} \hat{\psi}^{\dag}_\down(\br) \hat{\psi}_{\up}(\br)
    \ket{\phi_\up} 
= \rho_{\down\up;\up\down}^{*} \,.
\end{align}
\end{subequations}

Since the state $\ket{\phi_{\down}}$ has $N/2$ up 
and $N/2 +1$ down spins of the conduction electrons, the matrix element 
$\rho_{\up\up;\up\up}$ is given by the electron density with up spin
\begin{align}
\rho_{\up\up;\up\up}=\frac{N}{2V} =\frac{n}{2} \,,
\end{align}
where $V$ is the volume of the conduction electrons and $n$ is the electron 
density. Similarly we have $\rho_{\down\down;\down\down} ={n}/{2}\,.$
The matrix element $\rho_{\up\down;\up\down}$ is given by
\begin{align}
\rho_{\up\down;\up\down} = \frac{n}{2} + f(\br) \,,
\end{align}
where $f(\br)$ is defined by
\begin{align}
f(\br)\equiv \frac{1}{V{\cal N}} 
      \sum_{\bk\bk'>k_F} \Gamma_{\bk}\Gamma_{\bk'} e^{-i(\bk'-\bk)\cdot\br} \,. 
\end{align}
The system is isotropic, so $f(\br)$ depends only on the distance $r$.
We get the off-diagonal element $\rho_{\up\down;\down\up} = -f(r)\,.$
Thus $\rho^{(2)}$ is given by the sum of the fully mixed state and 
a spin-singlet state
\begin{subequations}
\label{Eq:two_density}
\begin{align}
\rho^{(2)} 
&= \frac{1}{2}
   \begin{bmatrix}
   \frac{n}{2} & 0 & 0 & 0 \\[6pt]
   0 & \frac{n}{2} + f(r) & -f(r)    & 0  \\[6pt]
   0 &-f(r)    & \frac{n}{2}+ f(r)   & 0  \\[6pt]
   0 &        0&                   0 &\frac{n}{2}
   \end{bmatrix} \,,\\
&= n\frac{\mathbb{I}}{4} + f(r)\ket{\Psi^{(-)}}\bra{\Psi^{(-)}}\,,
\end{align}
\end{subequations}
where  $\ket{\Psi^{(-)}}=\frac{1}{\sqrt{2}}(\ket{\up\down} -\ket{\down\up})$ and
$\mathbb{I}$ is the $4\times 4$ unit matrix. 
The meaning of Eq.~(\ref{Eq:two_density}) is that the first term is 
the uniform back ground and the second is the singlet state forming 
a Kondo screening cloud.

Since ${\rm Tr}\rho^{(2)}= n + f(r)$, let us introduce the normalized 
two-spin density matrix $\rho$ defined by $\rho^{(2)}= (n+f)\rho\,.$
Then we find that $\rho$ is given by the Werner state~\cite{Vedral03,Oh04}
\begin{align}
\rho = (1-p)\frac{\mathbb{I}}{4} + p \ket{\Psi^{(-)}}\bra{\Psi^{(-)}}\,,
\end{align}
where $p= f/(n+f)$. The impurity spin is entangled with one conduction 
electron spin at $\br$ if $p> 1/3$, that is, $f > n/2$.

Let us examine whether $\rho$ is entangled or not by calculating $f(r)$.
Since $\Gamma_{\bk}$ is real, we rewrite $f(r) = V{\tilde{f}}(r)^2/{\cal N}$
where the function $\tilde{f}(r)$ is defined by
\begin{align}
\tilde{f}(r) \equiv \frac{1}{V} \sum_{k>k_F} \Gamma_{\bk} e^{i\bk\cdot\br} \,.
\end{align}
After some calculations, $\tilde{f}(r)$ becomes the integral form 
\begin{align}
\tilde{f}(r) = \frac{N(0)}{k_Fr} \int_0^{D/E_F}\, 
       \frac{\sin\left[k_Fr\sqrt{1+ tE_B/E_F}\right]}{ t + 1} \,dt \,,
\end{align}
where $N(0)$ is the density of state at Fermi level and becomes
$N(0)= mk_F/2\pi^2\hbar^2 = 3n/4E_F$ for a free electron gas.
The normalization factor ${\cal N}$ is given by
\begin{subequations}
\begin{align}
{\cal N} &= \sum_{k>k_F}\Gamma_{\bk}^2 
          = \frac{VN(0)}{E_B} y(D,E_B, E_F) \,,
\end{align}
where the definite integral $y(D,E_B, E_F)$ is defined by 
\begin{align}
y(D,E_B, E_F)\equiv \int_0^{D/E_F} \frac{\sqrt{1 + t E_B/E_F}}{(1+t)^2}\,dt \,.
\end{align}
\end{subequations}
Thus we obtain
\begin{align}
f(r) =\frac{3n}{4} \frac{E_B}{E_F} \frac{f_N(r)^2}{y(D,E_B,E_F)} \,,
\end{align}
where $f_N(r)\equiv \tilde{f}(r)/N(0)$.
The condition of entanglement of $\rho$, $f>n/2$, becomes
\begin{align}
\frac{3}{2} \left(\frac{E_B}{E_F}\right) \frac{f_N(r)^2}{y(D,E_B,E_F)} > 1 \,.
\label{Eq:condition}
\end{align}

\begin{figure}[htbp]
\includegraphics[scale=1.0,angle=0]{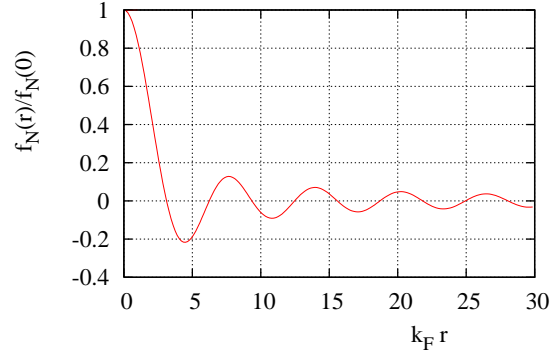}
\caption{$f_N(r)/f_N(0)$ as a function of $k_Fr$. 
The numerical values, $E_F,E_B$ and $D$ are taken in the text.}  
\label{Fig:fig2}
\end{figure}

For usual Kondo systems, the Kondo temperature $T_K$ ranges from few to 
hundreds Kelvin, $T_K \simeq  1\sim 300\, {\rm K}$. The Fermi temperature 
$T_F= E_F/k_B$ is about $T_F\simeq 10^4\,{\rm K}$. This implies 
$E_B/E_F \simeq 10^{-2} \sim 10^{-4}$. However, according to 
Eq.~(\ref{Eq:screening}), the small ratio of $E_B/E_F= k_BT_K/E_F$
indicates the Kondo screening length $\xi_K$ is much larger than
the Fermi wavelength $\lambda_F$. If the bandwidth $D$ is assumed to be 
$E_B \ll D \ll E_F$, we have $y(D,E_B,E_F) \simeq 1$. 
Also $f_N(r)\simeq {\cal O}(1)$ as shown in Fig.~(\ref{Fig:fig2}). 
Then $f/n$ could not be greater than the half, that is, 
Eq.~(\ref{Eq:condition}) is not satisfied. Therefore the impurity spin 
is not entangled with one conduction electron spin even within the Kondo 
screening length, i.e., $r < \xi_K$.
 
Fig.~(\ref{Fig:fig1}) shows the numerical calculation of $f_N(r)/f_N(0)$ 
as a function of $k_Fr$. $f_N(r)$ extends over the Kondo screening length 
$\xi_K$. We find that $f(r)$ is noting but the spin-spin correlation function 
between the impurity spin and the conduction spin at $\br$,
\begin{align}
\langle \sigma_{\rm im}^{z} \sigma^{z}(r) \rangle 
&= \rho_{\up\up;\up\up} + \rho_{\down\down;\down\down} 
 - \rho_{\up\down;\up\down} -\rho_{\down\up;\down\up} \nonumber \\[6pt]
&= -2f(r)\,.
\end{align}
Notice that $f(r)$ is qualitatively similar to the spatial spin-spin 
correlation function $\langle \mathbf{S}\cdot \mathbf{s}(\br)\rangle 
= \frac{3}{4}\langle\sigma_{\rm im}^{z} \sigma^{z}(r) \rangle$, which
was calculated by Chen {\it et al}~\cite{Chen87} and 
by Gubernatis {\it et al} for the Anderson model~\cite{Gubernatis87}.
It is interesting that $\rho^{(2)}$is given by a pseudo entangled 
state in liquid-state NMR~\cite{Nielsen00,Linden01}.
The Kondo screening cloud~\cite{Kouwenhoven01,Soerensen96} could be 
detected through the extra Knight shift experiment which measures $f(r)$. 

Our result shows that even if there is the spin-spin correlation between 
the impurity and a conduction electron in the Kondo screening cloud, 
entanglement between them vanishes. A simple explanation for this is 
as follows. Since many conduction electrons are coupled to a single impurity, 
the quantum correlation between the impurity and each conduction electron is 
very tiny so there is no entanglement between them. 

\paragraph{(iii) Entanglement between two conduction electron spins.}
It has been believed that conduction electrons within the Kondo screening 
cloud are mutually correlated because they have information on the same 
impurity~\cite{Kouwenhoven01}. We examine whether the impurity spin 
induces the non-classical correlation, i.e., entanglement between conduction 
electrons. 

Consider two conduction spins at $\br_1$ and $\br_2$ as shown 
in Fig~\ref{Fig:fig1}-(c). The density matrix $\rho_{\rm con}$ of all 
the conduction electrons is given by the mixture of $\ket{\phi_\up}$ 
and $\ket{\phi_\down}$
\begin{align}
\label{Eq:density_condution}
\rho_{\text{\rm con}} 
& = {\rm Tr}_{\rm im} \ket{\Psi_s} \bra{\Psi_s}
  = \frac{1}{2}\left(\, \ket{\phi_\up} \bra{\phi_\up}
                      + \ket{\phi_\down} \bra{\phi_\down} \,\right)\,.
\end{align}
From Eq.~(\ref{Eq:density_condution}), the density matrix $\tilde{\rho}^{(2)}$ 
of two conduction electron spins is written by
\begin{align}
&\tilde{\rho}_{\sigma_1',\sigma_2';\sigma_1,\sigma_2}^{(2)}(\br_1,\br_2) \nonumber\\
&= \frac{1}{2}{\rm Tr}
   \left[ 
    \hat{\psi}^\dag_{\sigma_2'}(\br_2)
    \hat{\psi}^{\dag}_{\sigma_1'}(\br_1)
    \hat{\psi}_{\sigma_1}(\br_1)\hat{\psi}_{\sigma_2}(\br_2)\,\rho_{\rm con}
   \right] \,.
\end{align}
After a lengthy calculation, we obtain
\begin{align}
\tilde{\rho}^{(2)} (\br_1,\br_2)
&= \rho^{(2)}_{\rm free}(r') + \Delta\rho(\br_1,\br_2)\,,
\end{align}
where $\br'\equiv \br_1-\br_2$ and $\rho^{(2)}_{\rm free}(r')$ is 
the density matrix of two electron spins of a free electron gas~\cite{Vedral03,Oh04}
\begin{align}
\rho^{(2)}_{\rm free}(r') = 
\frac{n^2}{8}
\begin{bmatrix}
1-g^2 & 0       & 0       & 0 \\
0        & 1       & -g^2 & 0 \\
0        & -g^2 & 1       & 0 \\
0        & 0       & 0       & 1 - g^2
\end{bmatrix}\,
\end{align}
with $g(r') \equiv\frac{2}{N}\sum_{k\le k_F} e^{i\bk\cdot\br'}$. 
The small change is 
\begin{align}
\Delta\rho(\br_1,\br_2) = 
\frac{n^2}{8}\,\frac{3E_B}{2E_F}
\begin{bmatrix}
a+b & 0   & 0   & 0\\
0   & a & b   & 0\\
0   & b   & a & 0\\
0   & 0   & 0   & a+b 
\end{bmatrix}\,,
\end{align}
where $a \equiv g(0)\,\left[\,f_N(r_1)^2 + f_N(r_2)^2\right]/2$ and
$b \equiv g(r')\,f_N(r_1)\,f_N(r_2)$.
Since ${E_B}/{E_F}$ is very small, we have $\rho^{(2)}(\br_1,\br_2) 
\approx \rho^{(2)}_{\rm free}(\br_1,\br_2)$. Of course, if $r_1,r_2 > \xi_K$, 
then one gets $\rho^{(2)}(\br_1,\br_2)= \rho^{(2)}_{\rm free}(\br_1,\br_2)$.

This result implies that the entanglement structure of 
the conduction electrons is little affected by the single impurity. 
In contrast to the traditional belief~\cite{Kouwenhoven01},
the mutual correlation between conduction electrons is mainly
due to the Pauli exclusion principle not due to the impurity. 
So their correlation length is the Fermi wavelength. The reason is that 
the number of conduction electrons are very large so they act as a reservoir.
Although $\ket{\phi_{\up}}$ is orthogonal to $\ket{\phi_{\down}}$,
$\ket{\phi_{\up}}$ differs from $\ket{\phi_{\down}}$ and from $\ket{F}$ 
by a single spin.

Before conclusion, we would like to mention an interesting experiment 
which reported the Ruderman-Kittel-Kasuya-Yosida (RKKY) interaction of two 
spin qubits on quantum dots mediated by an electron gas~\cite{Craig04,Glazman04}. 
The RKKY interaction may be used to couple spin qubits. However, our result 
implies that a spin qubit could be maximally entangled with the electron gas 
and become fully mixed if one does not deal with a quantum sate of 
a spin qubit and the electron gas as a whole. Our another study~\cite{Oh05b} 
on a two-spin boson model, similar to the two-impurity Kondo model, suggests 
that the electron gas could be used for an interaction mediator 
if the electron gas are separable from two spin qubits. 

Our analysis is based on Yosida's ground state which describes 
the Kondo physics in an simple but effective way. There are another ways 
to solve the problem: the variational ground state of the Anderson 
Hamiltonian~\cite{Varma76,Gunnarsson83}, the quantum Monte Carlo 
method~\cite{Gubernatis87}, the perturbative renormalization~\cite{Chen87}, 
and the density matrix renormalization group for a tight-binding 
model~\cite{Soerensen96}. 

In conclusion, we have studied entanglement of the impurity spin and 
the conduction electron spins in the Kondo model based on Yosida's ground 
state. First, it has been shown that the impurity spin is maximally entangled 
with all the conduction electrons. Second, the two-spin density matrix of 
the impurity spin and one conduction electron spin at $\br$ is given 
by a Werner state. It has been found that the impurity spin is not entangled 
with a conduction electron spin within the Kondo screening cloud 
even though there exists the spin-spin correlation between them. 
Third, the entanglement structure of the conduction electrons is little affected 
by the impurity in contrast to the strong effect on electrical resistance.

\acknowledgments
We thank Prof. K. Yamada of University of Tokyo for helpful comments.
This work was supported by Korean Research Foundation Grant 
(KRF-2004-041-C00089). J.K. was supported by a Grant (TRQCQ) from the
Ministry of Science and Technology of Korea.

\end{document}